\documentclass[%
preprint,
amsmath,amssymb,
aps,
]{revtex4-2}

\usepackage{graphicx}% Include figure files
\usepackage{dcolumn}% Align table columns on decimal point
\usepackage{bm}% bold math
\def\dalam{
	\vbox{\hsize 0.5 em \hrule\hbox to 0.5 em{\vrule width 0.5 pt height 0.5 em
			\hfill\vrule width 0.5 pt height 0.5 em} \hrule}  }

\allowdisplaybreaks
\begin{document}%
	%%%%%%%%%%%%%%%%%
	
	\title{Dilaton photoproduction in a magnetic dipole field of pulsars and magnetars}
	\author{M.O. Astashenkov}
	\affiliation{Physics department Moscow State University, Moscow 119991, Russia}
	\begin{abstract}
		According to Einstein-Maxwell-Dilaton theory, the dilaton field $\psi$ can be produced by electromagnetic fields with non-zero Maxwell invariant. So electromagnetic wave propagating in an external electromagnetic field is a typical source of dilaton radiation.
		For study dilaton photoproduction in astrophysical conditions it's interesting to consider plane elliptically polarized electromagnetic wave propagating in the electromagnetic field of magnetic dipole ${\bf m}$ of pulsars and magnetars. 
		The dilation field equation is solved in case $|\psi| \ll 1$. 
		The angular distribution dilaton radiation is studied in every point of space. It's shown that spectral composition of dilatons is similar to spectral composition of plane electromagnetic wave. Amount of dilaton energy radiated in time and all directions is greatest in condition $(B_1^2-B_2^2)(m_x^2-m_y^2)\geq 0,$ where $B_1$ and $B_2$ are electromagnetic wave amplitudes along the axes of polarization ellipse. 
		This condition is valid for many neutron star systems.
	\end{abstract}



\maketitle	
\newpage	

	\section{Introduction}
	
	Nowadays in scientific literature there are few theories for  Goldstone bosons:arions \cite{axion-1, axion-2, axion-3, axion-4, axion-5}, axions \cite{axion-6, arions-from-pulsar, axion-8} and dilatons \cite{, axion-9, axion-10, axion-11, axion-12, axion-13, axion-14}. 
	One of the main sources of these particles is electromagnetic fields and waves. Equations for these bosons in the classic limit have similar form.
	In particular, corresponding to string theory system of interacting Maxwell (U(1) gauge) and scalar (dilaton) fields  design, so called Einstein-Maxwell-Dilaton theory.
	
	According to \cite{axion-15}, the action of Einstein-Maxwell-Dilaton theory can be written as
	\begin{equation} \label{S}
		S=\int d^4x\Big\{a_0(\partial\Psi)^2+a_1e^{-2{\cal K}\Psi}F^{nm}F_{nm}\Big\},
	\end{equation}
	where $\Psi$ is a dilaton field, $a_0, a_1$ and $\cal K$ are gauge constants and $F_{nm}$ is Maxwell tensor. 
	
	In string theory ${\cal K} = 1$, the five-dimensional Kaluza-Klein theory 
	results in the value ${\cal K} = \sqrt{3}$;  In this work  the constant ${\cal K}$ is arbitrary. 
	
	The field equations in Minkowski spacetime obtained from action (\ref{S}) have the form:
	\begin{equation} \label{field_eq1}
		\dalam \Psi={a_1{\cal K}\over a_0}e^{-2{\cal K}\Psi} F_{nm}F^{nm}
		={2a_1{\cal K}\over a_0} e^{-2{\cal K}\Psi}\big[{\bf B^2}-{\bf E^2}\big],
	\end{equation}
	\begin{equation}	
		\label{field_eq2}
		{\partial \over \partial x^n}\Big[e^{-2{\cal K}\Psi}F^{nm}\Big]=0\;,
	\end{equation}
	where $\bf E$ is the electric field strength and $\bf B$ is  magnetic induction.
	
	As dilaton field hasn't been discovered, one can assume that
	dilaton field is weak in Solar system and $|\psi|\ll 1$. In this case equation (\ref{field_eq2}) can be expressed as
	
	\begin{equation} \label{dilaton_eq}
		\dalam \Psi= \frac{2a_1{\cal K}}{a_0} \big[{\bf B^2}-{\bf E^2}\big]\;.
	\end{equation}
	According to equations (\ref{field_eq2}) - (\ref{dilaton_eq}), 
	the invariant ${\bf B^2} - {\bf E^2}$ is the source of the dilaton field. Beside this invariant is equal to zero for wave zone of every electromagnetic wave, dilaton photoproduction is possible from near zone or in area where there is a superposition of electromagnetic fields with non-zero invariant ${\bf B^2} - {\bf E^2}$.
	
	\section{Basic equation and its solution}
	Consider a plane elliptically polarized electromagnetic wave with frequency $\omega$ propagating among axis $z$. 
	Fields {\bf E} and {\bf B} of electromagnetic wave have the following form:
	\begin{eqnarray} \label{wave}
			{\bf B}={\bf e_x} B_1\cos(kz-\omega t)-{\bf e_y} B_2\sin(kz-\omega t), \\* \nonumber
			{\bf E}=-{\bf e_x} B_2\sin(kz-\omega t)-{\bf e_y} B_1\cos(kz-\omega t),
	\end{eqnarray}   
	where $k = \omega/c$, $B_1$ and $B_2$ are the electromagnetic wave amplitudes among principal axes of the polarization ellipse.
	
	For example, if $B_1 = B_2$ then the wave ~(\ref{wave}) has circle polarization; if $B_1 = 0$ or $B_2 = 0$ then the wave ~(\ref{wave}) is linear polarized. In other cases wave has elliptical polarization.
	
	Assume that there is a neutron star with radius $R_S$ rotating around magnetic dipole momentum ${\bf m}$ in the origin of the axis. The magnetic induction ${\bf B}$ of the neuron star for $r>R_S$ has the form:
	\begin{equation} \label{m_dip}
		{\bf B}=\frac{3({\bf m}\cdot {\bf r}){\bf r} - {\bf m} r^2}{r^5}\; ,
	\end{equation}
	For now a few hundred neutron stars have been discovered \cite{pulsar-cat, magnetar-cat} whose rotating axis doesn't coincide with the axis of magnetic dipole momentum. Such stars radiate electromagnetic waves in magnetic dipole approximations (pulsars and magnetars). But there must be neutron stars whose dipole momentum axis coincides with the rotation axis. In this case there is no electromagnetic radiation and their magnetic field ~(\ref{m_dip}) must be static.
	
	Substituting superposition of electromagnetic fields ~(\ref{wave}) and ~(\ref{m_dip}) to expression ~(\ref{dilaton_eq}) and discarding the time-independent terms, one can obtain 
	\begin{eqnarray} \label{Re_main_equation}
			\dalam \Psi=\frac{4a_1{\cal K}}{a_0r^5} \Big\{3({\bf m}\cdot {\bf r})
			\big[xB_1\cos(kz-\omega t)-y B_2\sin(kz-\omega t)\big]\\* \nonumber
			-r^2\big[m_xB_1\cos(kz-\omega t)-m_y B_2\sin(kz-\omega t)\big]\Big\}.
	\end{eqnarray}
	
	Exact solution of the Eq.~(\ref{Re_main_equation}) is found to be
	\begin{eqnarray} \label{Re_solution}
			\Psi= \frac{2a_1{\cal K}}{ka_0}
			\Big\{k\Big[\frac{(m_xx+m_yy)}{r^2(r-z)}
			-\frac{m_z}{r^2}\Big]
			\times\Big[B_1x\cos k(r-ct) 
			-B_2y\sin k(r-ct)\Big] \\* \nonumber
			-\Big[\frac{(m_xx+m_yy)(2r-z)}{r^3(r-z)^2} 
			-\frac{m_z}{r^3}\Big] 
			\times\Big[B_1x[\sin k(r-ct)-\sin k(z-ct)] \\* \nonumber
			+B_2y[\cos k(r-ct)-\cos k(z-ct)]\;\Big] +\frac{1}{r(r-z)}\Big[B_1m_x[\sin k(r-ct) \\*\nonumber
			-\sin k(z-ct)]
			+B_2m_y[\cos k(r-ct)-\cos k(z-ct)]\;\Big]\Big\}.
	\end{eqnarray}

	It follows from this expression that spectral composition of dilatons coincides with the spectral composition of plane electromagnetic wave.
	
	Rewrite expression ~(\ref{Re_solution}) in spherical coordinates:
	\begin{eqnarray} \label{psi_spherical}
			\Psi= {2a_1{\cal K}\over ka_0}\Big\{k\Big[\frac{[m_x\cos\phi+m_y\sin\phi]\sin\theta}{r(1-\cos\theta)}
			-{m_z\over r}\Big]\Big[B_1\cos\phi\cos k(r-ct)\\* \nonumber
			-B_2\sin\phi\sin k(r-ct)\Big]\sin\theta-\Big[\frac{[m_x\cos\phi+m_y\sin\phi](2-\cos\theta)
				\sin\theta}{r^2(1-\cos\theta)^2}
			-{m_z\over r^2}\Big] \\* \nonumber
			\times\Big[B_1\cos\phi[\sin k(r-ct)-\sin k(r\cos\theta-ct)]
			+B_2\sin\phi[\cos k(r-ct)\\* \nonumber
			-\cos k(r\cos\theta-ct)]\Big]\sin\theta+\frac{1}{ r^2(1-\cos\theta)}
			\Big[B_1m_x[\sin k(r-ct)\\* \nonumber
			-\sin k(r\cos\theta-ct)]+B_2m_y[\cos k(r-ct)-\cos k(r\cos\theta-ct)]\Big]\Big\}.
	\end{eqnarray}
	It should be noted that in limit $\theta \to 0$ dilaton field has no singularity
	\begin{equation}
		\lim_{\theta\to 0}\Psi={2a_1{\cal K}\over a_0}\Big\{B_1m_x\cos k(r-ct)-B_2m_y\sin k(r-ct)\Big\}.
	\end{equation}
	Thus, dilaton field has finite value everywhere out of neutron star.
	Using the expression ~(\ref{psi_spherical}) one can study angular distribution of the dilaton radiation.
	
	\section{Angular distribution of the dilaton radiation}
	
	By definition \cite{arions-from-pulsar, axion-18}, the amount of energy $dI$ emitted by the source per unit time through the solid angle $d\Omega$ is given by the formula:
	\begin{equation}
		\frac{dI}{d\Omega} = \lim_{r\to \infty} r({\bf W} \cdot {\bf r}) \;,
	\end{equation}
	where ${\bf W}$ is the energy flux density vector associated with the components of the stress-energy tensor $T^{ik}$ by the relations: $W^{\alpha} = T^{0\alpha}$.
	
	For free dilatonic field  the stress-energy tensor  $T^{ik}$ has the form:
	\begin{equation}
		T^{ik}=2a_0g^{in}g^{km} \big\{{\partial \Psi\over \partial x^n}
		{\partial \Psi\over \partial x^m}
		-{1\over 2}g^{ik}{\partial \Psi\over \partial x^n}
		{\partial \Psi\over \partial x^m} g^{nm}\big\}.
	\end{equation}
	It follows that for $a_0>0$ dilaton energy density is positive for every distribution of electromagnetic fields.
	\begin{equation}
		T^{00}=a_0\Big\{{1\over c^2}\left({\partial \Psi\over \partial t}\right)^2
		+\left({\partial \Psi\over \partial x}\right)^2+\left({\partial \Psi\over \partial y}\right)^2
		+\left({\partial \Psi\over \partial z}\right)^2\Big\}\geq0.
	\end{equation}
	
	Angular distribution of dilaton radiation produced by elliptically polarized electromagnetic wave ~(\ref{wave}) propagating in magnetic field of neutron star ~(\ref{m_dip}) can be calculated by formula:
	\begin{equation} \label{dI_dOmega_formula}
		{dI\over d\Omega}=r\big({\bf r\ W}\big)= -2a_0r
		(\vec r\ \vec \nabla \Psi){\partial \Psi\over\partial t},
	\end{equation}
	In the expression above it is taken into account that the dilaton field $\psi$ has a special point in $\theta = 0$.
	
	Substituting ~(\ref{Re_solution}) to ~(\ref{dI_dOmega_formula}) one can obtain angular distribution averaged over the period of electromagnetic wave $T = 2\pi/\omega$:
	
	\begin{eqnarray}
	\overline{\frac{dI}{d\Omega}}=\frac{4a_1^2{\cal K}^2c}{ a_0r^2}\Big\{
		%%%%%%%%%%%%%%%%%%%%%%%%%%%%%%%%%%%%%%%%%%%%%%%%%%%%%%%%%%%%%%
	\;4k^2r^2(B_1^2\cos^2\phi  + B_2^2\sin^2\phi )(m_x\cos\phi +m_y\sin\phi )^2 
	+2kr\big(\\ \nonumber  B_1B_2m_y\cos\phi  
	- B_1B_2m_x\sin\phi 
	- 2krB_1^2m_z\sin\theta\cos^2\phi 
	- 2krB_2^2m_z \sin\theta \\ \nonumber 
	\times\sin^2\phi 
	\;\big) 
	(m_x\cos\phi  +m_y\sin\phi ) 
	+krB_1B_2m_z\sin\theta 
	( m_x\sin\phi  - m_y\cos\phi )  \\[1ex] \nonumber
	%%%%%%%%%%%%%%%%%%%%%%%%%%%%%%%%%%%%%%%%%%%%%%%%%%%%%%%%%%%%%%%
	+\;(1-\cos\theta) \times\Big[
	2k^2r^2m_z^2( B_1^2\cos^2\phi  + B_2^2\sin^2\phi ) 
	-4k^2r^2(B_1^2\cos^2\phi    +B_2^2\sin^2\phi ) \\ \nonumber
	\times(m_x\cos\phi +m_y\sin\phi)^2 
	+kr\big( 2krB_1^2m_z\sin\theta\cos^2\phi 
	+ 2krB_2^2m_z\sin\theta\sin^2\phi  \\ \nonumber
	+ B_1B_2m_x\sin\phi 
	- B_1B_2m_y\cos\phi 
	\big) (m_x\cos\phi +m_y\sin\phi ) 
	\Big]\\ \nonumber
%%%%%%%%%%%%%%%%%%%%%%%%%%%%%%%%%%%%%%%%%%%%%%%%%%%%%%%%%%%%%%%
	+(1-\cos\theta)^2\Big[k^2r^2(m_x\cos\phi+m_y\sin\phi )^2
	-k^2r^2m_z^2 
	+\big(\;3(m_x\cos\phi  +m_y\sin\phi )^2 \\ \nonumber 
	- 3m_z^2 
	-2m_z\sin\theta (m_x\cos\phi +m_y\sin\phi )
	\;\big) [1-\cos kr(1-\cos\theta)] 
	+\big(\;4krm_z^2 
	+2kr\\ \nonumber \times m_z\sin\theta 
	(m_x\cos\phi +m_y\sin\phi ) 
	-5kr(m_x\cos\phi +m_y\sin\phi )^2\;\big)\sin kr(1-\cos\theta)\Big] \\ \nonumber
	\times(B_1^2\cos^2\phi  + B_2^2\sin^2\phi  ) 
	%%%%%%%%%%%%%%%%%%%%%%%%%%%%%%%%%%%%%%%%%%%%%%%%%%%%%%%%%%%
	\;+\;(1-\cos\theta)^3\times\Big[\big(m_z^2-(m_x\cos\phi +m_y\sin\phi )^2\big)\\ \nonumber
	\times[1-\cos kr(1-\cos\theta )]
	+kr(\;(m_x\cos\phi +m_y\sin\phi )^2-m_z^2\;)\sin kr(1-\cos \theta)\Big] \\ \nonumber
	\times\Big(B_1^2\cos^2\phi  +B_2^2\sin^2\phi\Big)
	%%%%%%%%%%%%%%%%%%%%%%%%%%%%%%%%%%%%%%%%%%%%%%%%%%%%%%%%%
	+\;\frac{[1-\cos kr(1-\cos\theta)]}{(1-\cos\theta)^2} \times
	\Big[8(B_1^2\cos^2\phi +B_2^2\sin^2\phi)\\ \nonumber
	\times(m_x\cos\phi+m_y\sin\phi)^2 
	+2(B_1^2m_x^2 + B_2^2m_y^2)
	- 8(B_1^2m_x\cos\phi  +B_2^2m_y\sin\phi )\\ \nonumber
	\times(m_x\cos\phi +m_y\sin\phi )\Big]
	%%%%%%%%%%%%%%%%%%%%%%%%%%%%%%%%%%%%%%%%%%%%%%%%%%%%%
	+\;\frac{[1-\cos kr(1-\cos\theta)]}{(1-\cos\theta)} \times
	\Big[2\big(B_1^2m_x\cos\phi + B_2^2m_y\sin\phi \\ \nonumber
	 -3B_1^2m_z\sin\theta\cos^2\phi  -3B_2^2m_z\sin\theta\sin^2\phi
	+2krB_1B_2m_y\cos\phi  
	- 2krB_1B_2m_x\sin\phi \big)\\ \nonumber
	\times(m_x\cos\phi +m_y\sin\phi )
	+3B_1^2m_xm_z\sin\theta\cos\phi  +3B_2^2m_ym_z\sin\phi \sin\theta   \\ \nonumber
	- 2krB_1B_2m_ym_z\sin\theta\cos\phi   
	+ 2krB_1B_2m_xm_z\sin\phi \sin\theta    
	- B_1^2m_x^2  - B_2^2m_y^2\Big]\\[2ex] \nonumber
	%%%%%%%%%%%%%%%%%%%%%%%%%%%%%%%%%%%%%%%%%%%%%%%%%%%%%%%%%%%%%%%
	+\;\frac{\sin(kr(1-\cos\theta))}{(1-\cos\theta)}\times
	\Big[m_z\sin\theta \big(B_1B_2m_y\cos\phi 
	- B_1B_2m_x\sin\phi - 2krB_1^2m_x\cos\phi   \\ \nonumber
	- 2krB_2^2m_y\sin\phi\big) 
	-8kr(B_1^2\cos^2\phi  +
	B_2^2\sin^2\phi )(m_x\cos\phi +m_y\sin\phi )^2 \\ \nonumber
	+2\big(2krB_1^2m_x\cos\phi + 2krB_2^2m_y\sin\phi  - B_1B_2m_y\cos\phi  +B_1B_2m_x\sin\phi\\ \nonumber
	+ 2krB_1^2m_z\sin\theta\cos^2\phi   + 2krB_2^2m_z \sin\theta\sin^2\phi   \big)(m_x\cos\phi +m_y\sin\phi )\Big]\\[2ex] \nonumber
	%%%%%%%%%%%%%%%%%%%%%%%%%%%%%%%%%%%%%%%%%%%%%%%%%%%%%%%%%%%%%%%
	+\;[1-\cos kr(1-\cos\theta )]\times\Big[\big(5B_1^2m_x\cos\phi + 5B_2^2m_y\sin\phi  - 4krB_1B_2m_y\cos\phi \\ \nonumber
	+ 4krB_1B_2m_x\sin\phi + 3B_1^2m_z\sin\theta\cos^2\phi
	+ 3B_2^2m_z\sin\theta\sin^2\phi   
	\big)(m_x\cos\phi\\ \nonumber
	 +m_y\sin\phi )
	-10(B_1^2\cos^2\phi  +B_2^2\sin^2\phi )(m_x\cos\phi +m_y\sin\phi )^2 
	+m_z\sin\theta \\ \nonumber
	\times \big(krB_1B_2m_y\cos\phi     
	- krB_1B_2m_x\sin\phi - 2B_1^2m_x\cos\phi 
	- 2B_2^2m_y\sin\phi \big)\Big]\\[1ex] \nonumber
	%%%%%%%%%%%%%%%%%%%%%%%%%%%%%%%%%%%%%%%%%%%%%%%%%%%%%%%%%%%%%%%
	+\;\sin kr(1-\cos\theta )\times\Big[4kr( B_1^2\cos^2\phi  + B_2^2\sin^2\phi )(m_x\cos\phi +m_y\sin\phi )^2\\ \nonumber
	+\big( 
	4krB_1^2m_z\sin\theta\cos^2\phi  
	+ 4krB_2^2m_z\sin\theta\sin^2\phi   
	- 4krB_1^2m_x\cos\phi  \\ \nonumber
	- 4krB_2^2m_y\sin\phi
	+ B_1B_2m_y\cos\phi 
	- B_1B_2m_x\sin\phi   \big)(m_x\cos\phi \\ \nonumber
	+m_y\sin\phi ) 
	+krm_z\sin\theta (B_1^2m_x\cos\phi  + B_2^2m_y\sin\phi )\Big]
	\\[1ex] \nonumber
	%%%%%%%%%%%%%%%%%%%%%%%%%%%%%%%%%%%%%%%%%%%%%%%%%%%%%%%%%%%%%%%
	+\;(1-\cos\theta)[1-\cos kr(1-\cos\theta)]\times\Big[2m_z^2(B_1^2\cos^2\phi  + B_2^2\sin^2\phi) \\ \nonumber  
	+2(B_1^2\cos^2\phi 
	+ B_2^2\sin^2\phi ) (m_x\cos\phi +m_y\sin\phi )^2  
	+\big(  
	4B_1^2m_z\sin\theta\cos^2\phi \\ \nonumber
	+ 4B_2^2m_z\sin\theta\sin^2\phi 
	- 2B_1^2m_x\cos\phi  	
	- 2B_2^2m_y\sin\phi 
	+ krB_1B_2m_y\cos\phi \\ \nonumber
	-  krB_1B_2m_x\sin\phi 
	\big)(m_x\cos\phi +m_y\sin\phi )
	\Big]
	%%%%%%%%%%%%%%%%%%%%%%%%%%%%%%%%%%%%%%%%%%%%%%%%%%%%%%%%%%%%%
	+\;(1-\cos\theta )\sin kr(1-\cos\theta)\\ \nonumber
	\times
	\Big[6kr(B_1^2\cos^2\phi  + B_2^2\sin^2\phi )
	(m_x\cos\phi +m_y\sin\phi )^2 
	-4krm_z^2(B_1^2\cos^2\phi  \\ \nonumber
	+ B_2^2\sin^2\phi ) 
	+kr(B_1^2m_x\cos\phi 
	+ B_2^2m_y\sin\phi 
	- 7B_1^2m_z\sin\theta\cos^2\phi \\ \nonumber
	- 7B_2^2m_z\sin\theta\sin^2\phi  
	)(m_x\cos\phi +m_y\sin\phi )
	\Big]\Big\}.
\end{eqnarray}	

	For dilaton radiation forward ($\theta = 0$) one can obtain:
	\begin{equation} 
		\overline {\frac{dI}{d\Omega}}(\theta=0)={4a_1^2{\cal K}^2\omega^2\over a_0c}
		\Big(B_1^2m_x^2+B_2^2m_y^2\Big). 
	\end{equation}
	There is no dilaton radiation backward $dI/d\Omega(\theta = \pi) = 0$.
	
	Integrating over angles $\theta$, $\phi$ one can obtain amount of dilaton energy $\overline{I}$ averaged over the period of electromagnetic wave in main asymptotic approximation ($r \to \infty$) radiating per unit of time in all directions:
	\begin{equation} \label{full_intensity}
		\overline {I}=\frac{8\pi a_1ck^2{\cal K}}{3a_0}
		\Big\{3(B_1^2m_x^2+ B_2^2m_y^2) + 2(B_1^2+B_2^2)m_z^2 + B_1^2m_y^2+ B_2^2m_x^2\Big\}\;.
	\end{equation}
	It follows from expression ~(\ref{full_intensity}) that amount of dilaton energy $\overline{I}$ per unit of time is the greatest in condition $(B_1^2 - B_2^2)(m_x^2 - m_y^2) > 0$.
	
	\section{conclusion}
	As it was shown, a plane elliptically polarized electromagnetic wave propagating in an electromagnetic field of magnetic dipole produces a dilaton wave, whose amplitude has a special point in $\theta = 0$. But dilaton $\psi$ has a finite value at this point. That's why the dilaton field has finite value in an area where $r>R_S$.
	
	The dilaton radiation forward ($\theta = 0$) has the form:
	\begin{equation}
		\overline {\frac{dI}{d\Omega}}(\theta=0)={4a_1^2{\cal K}^2\omega^2\over a_0c}
		\Big(B_1^2m_x^2+B_2^2m_y^2\Big).
	\end{equation}
	There is no dilaton radiation backward $dI/d\Omega(\theta = \pi) = 0$.
	
	Simple analysis shows that the amount of dilaton energy $\overline{I}$ per unit of time averaged over the period of electromagnetic wave is the greatest in condition $(B_1^2 - B_2^2)(m_x^2 - m_y^2) > 0$. 
	This condition is valid for many superpositions of electromagnetic waves and magnetic dipole momentums of neuron stars.

	\section{Acknowledgements}
	The research was carried out within the framework of the scientific program of 
	the National Center for Physics and Mathematics, the project "Particle Physics and Cosmology".

	%%%%%%%%%%%%%%%%

\begin{thebibliography}{100}
		
		\bibitem{axion-1}
		G. Raffelt and D. Dearbom, {\it Bounds on hadronic axions from stellar evolution}, Phys. Rev. D, {\bf 36} 2211 (1987).
		
		\bibitem{axion-2}	
		G. Raffelt, {\it Astrophysical methods to constrain axions and other novel	particle phenomena},
		Phys. Rep., {\bf 198}, 1 (1990).
		
		\bibitem{axion-3}	
		G.Raffelt and L.Stodolsky, {\it Mixing of the photon with low-mass particles}, Phys. Rev. D {\bf 37}, 1237 (1988).       
		
		\bibitem{axion-4}
		Lescano, E., Miron-Granese, N.  {\it Double field theory with matter and its cosmological application}, Phys. Rev.D, {\bf 107}, 046016 (2023).
		
		\bibitem{axion-5}	
		D.Banerjee et al. (NA64 Collaboration), Search for             
		axionlike and scalar particles with the NA64 experiment,
		Phys. Rev. Lett. {\bf 125}, 081801 (2020).
		
		\bibitem{axion-6}	
		A.A.Anselm and N.G.Uraltsev, A second massless axion?, Phys. Lett. B {\bf 114},39 (1982).
		
		\bibitem{arions-from-pulsar}	
		V.I. Denisov, B.D. Garmaev, and I.P. Denisova, {\it Radiation of arions by electromagnetic field of rotating magnetic dipole},
		Phys. Rev. D, {\bf 104}, 055018 (2021). %https://doi.org/10.1103/PhysRevD.104.055018\\
		
		\bibitem{axion-8}
		I.P.Denisova, {\it On a mathematical problem in the theory of Goldstone Bosons},
		Russian Mathematics,  {\bf 64}, 73 (2020).
		
		\bibitem{axion-9}	
		H. C. L.Junior, et al.  {\it Einstein-Maxwell-dilaton neutral black holes in strong magnetic fields: Topological charge, shadows, and lensing}. Phys. Rev. D, {\bf 105}, 064070 (2022).
		
		\bibitem{axion-10}	
		Ahmed, A., Najjari, S. {\it Ultraviolet freeze-in dark matter through the dilaton portal}, Phys. Rev. D, {\bf 107}, 055020 (2023).
		
		\bibitem{axion-11}	
		Zhang, C. Y., Liu, P., Liu, Y., Niu, C., Wang, B. {\it Dynamical scalarization in Einstein-Maxwell-dilaton theory}, Phys. Rev. D, {\bf 105},	024073 (2022).
		
		\bibitem{axion-12}
		I.P.Denisova, {\it Dilatons Generation while Plane Electromagnetic Wave Propagates in Coulomb Field}, Gravitation and Cosmology, {\bf 27},  392 (2021).
		
		\bibitem{axion-13}	
		Shahkarami L. {\it Magnetized Einstein-Maxwell-dilaton model under an external electric field}, The Eur. Phys. J. C. {\bf 82},  1 (2022).
		
		\bibitem{axion-14}
		A. N. Malybayev, K. A. Boshkayev, V. D. Ivashchuk,
		{\it Quasinormal modes in the field of a dyon-like dilatonic Black hole},
		Eur. Phys. J. C {\bf 81} (2021) 475. %https://doi.org/10.1140/epjc/s10052-021-09252-z\\
		
		\bibitem{axion-15}	
		V. I. Denisov, I. P. Denisova and E. T. Einiev, {\it The investigation of low-frequency 
			dilaton generation},  Eur. Phys. J. C {\bf 82} 311 (2022).
		\bibitem{pulsar-cat}	
		R.N.Manchester, G.B.Hobbs, A.Teoh, M.Hobbs,  
		{\it The Australia telescope national facility pulsar catalogue}, 
		The Astron. J.,  {\bf 129}, 1993 (2005).
		
		\bibitem{magnetar-cat}	
		S.A.Olausen and V.M.Kaspi, {\it The McGill magnetar catalog},
		Astrophys. J. Suppl. {\bf 212}, 6 (2014).
		
		\bibitem{axion-18}	
		L.D.Landau, E.M.Lifshitz, {\it The classical theory of fields},
		(Butterworth-Heinemann, 1975).
		
	\end{thebibliography}
\end{document}